# The Ap 2-1 nebula and the surrounding molecular cloud G35.2-0.74: an active star forming region

S. Paron[1]* and W. Weidmann[2]

[1] *Instituto de Astronomía y Física del Espacio (CONICET-UBA), CC 67, Suc. 28, 1428 Buenos Aires, Argentina*
[2] *Observatorio Astronómico Córdoba, Universidad Nacional de Córdoba, Argentina*

**ABSTRACT**

Using data from large-scale surveys: 2MASS, GLIMPSE, MIPSGAL, VGPS, GRS, and IPHAS, we performed a multiwavelength study of the ISM in a region of about $20' \times 20'$ towards the molecular cloud G35.2-0.74. Additionally, the Ap 2-1 nebula, that is seen in projection over the molecular cloud, was studied using optical data obtained with the 2.15 m telescope at CASLEO, Argentina. From the HI absorption study we estimate a distance of $\sim 2$ kpc for Ap 2-1 confirming that the nebula is embedded in the south portion of the molecular cloud G35.2-0.74. Performing a photometric study and analysing the spectral energy distributions of the sources likely embedded in the cloud, we confirm that this region is very active in star formation, mainly towards the north, where we discover a cluster of young stellar objects. From the $H\alpha$ and [NII] lines we obtain a radial velocity of $v_{LSR} \sim 31$ km s$^{-1}$ for the Ap 2-1 nebula, in coincidence with the velocity of the molecular cloud. Finally, we conclude that Ap 2-1 is an HII region probably excited by an early B-type star.

**Key words:** ISM: clouds - (ISM:) HII regions - Stars: formation

## 1 INTRODUCTION

Ap 2-1 (PN G035.1-00.7; R.A.= $18^h58^m10.48^s$, DEC. = + $01°36'57''.2$, J2000.0) was discovered by Apriamasvili (1962) at Monte Kanobil. This nebula is a peculiar object poorly studied. At optical wavelengths, Ap 2-1 presents an uniform brightness and a roughly elliptical shape in a position angle of 146°. The optical angular diameter of Ap 2-1 reported in the literature ranges from $27''$ (Zijlstra et al. 1990) to $36''$ (Milne & Aller 1982). A bright central source is located at the nucleus with a magnitude of V = 15.16 (Tylenda et al. 1991). Although Ap 2-1 was classified as a planetary nebula (PN) (Acker et al. 1992; Kohoutek 2001), there are evidence suggesting that it could be a compact HII region (Glushkov et al. 1975; Acker et al. 1987; Zijlstra et al. 1989; Soker 1997). In fact, Ap 2-1 was cataloged as a compact HII region by Giveon et al. (2005a,b). Moreover, Little et al. (1985) detected emission of NH$_3$ towards Ap 2-1, which is typically observed in high density ambients such as star forming regions, ultracompact or compact HII regions.

Ap 2-1 is seen in projection over the south portion of the molecular cloud G35.2-0.74, called G35.2S by Dent et al. (1984). The north portion of this cloud, G35.2N (centered at R.A.= $18^h58^m13^s$, DEC. = + $01°40'26''$, J2000.0), is an active star forming region with outflow activity (Dent et al. 1984; Gibb et al. 2003; Birks et al. 2006). In fact, this region was cataloged as EGO G35.20-0.74 by Cyganowski et al. (2008). An EGO, "extended green object", is a source with extended *Spitzer*-IRAC 4.5 $\mu$m emission, which is usually presented in green colour. According to the authors, an EGO is a massive young stellar object (MYSO) driving outflows. The molecular cloud G35.2-0.74 is at the distance of about 2 kpc (Birks et al. 2006). Recently Phillips & Ramos-Larios (2008) studied the region in the mid infrared emission, they suggest that Ap 2-1 is apparently enclosed within a very much larger HII region.

In this work we study the ISM in a large portion around Ap 2-1 nebula, we look for star formation signatures in the molecular cloud G35.2-0.74, and finally we present an optical analysis of Ap 2-1.

## 2 DATA FROM LARGE-SCALE SURVEYS

In order to study the surrounding ISM of Ap 2-1, we analysed molecular and IR data from large-scale surveys. We used data from the Galactic Ring Survey (GRS), the Galactic Legacy Infrared Mid-Plane Survey Extraordinaire (GLIMPSE), MIPSGAL and Two Micron All Sky Survey (2MASS). The GRS is being performed by the Boston University and the Five College Radio Astronomy Observatory (FCRAO). The survey maps the Galactic Ring in the $^{13}$CO J=1–0 line with an angular and spectral resolution of 46″ and 0.2 km s$^{-1}$, respectively (see Jackson et al. 2006). The observations were performed in both position-switching and On-The-Fly mapping modes, achieving an angular sampling of 22″. GLIMPSE is a mid infrared survey of the inner Galaxy performed using the *Spitzer Space Telescope*. We used the mosaiced images from GLIMPSE and the GLIMPSE Point-Source Catalog (GPSC) in the *Spitzer*-IRAC (3.6, 4.5, 5.8 and 8 $\mu$m). IRAC has an angular resolution between 1″.5 and 1″.9 (see Fazio et al. 2004 and Werner et al. 2004). MIPSGAL is a survey of the same region as GLIMPSE, using MIPS instrument (24 and 70 $\mu$m) onboard *Spitzer*. The MIPSGAL resolution at 24 $\mu$m is 6″. 2MASS is a joint project of the University of Massachusetts and the Infrared Processing and Analysis Center/California Institute of Technology, funded by the National Aeronautics and Space Administration and the National Science Foundation. We used the 2MASS All-Sky Point Source Catalog (PSC). Additionally, we used HI and $H\alpha$ data extracted from the VLA Galactic Plane Survey (VGPS; Stil et al. 2006) and from the INT Photometric $H\alpha$ Survey of the Northern Galactic Plane (IPHAS, Drew et al. 2005), respectively.

---

* E-mail: sparon@iafe.uba.ar

## 3 THE ISM AROUND AP 2-1

### 3.1 Distance

Taking into account that Ap 2-1 presents radio continuum emission at 20 cm in the VGPS, we use HI data extracted from the same survey to perform an absorption study towards the nebula to estimate its distance. The angular resolution and sensitivity of the data are 1′ and 2 K, respectively. We are studying a region in the first galactic quadrant, thus we have to take into account the distance ambiguity that exists when using radial velocities and a galactic rotation curve to assign distances to sources. Figure 1 shows the HI spectra: red is the HI emission obtained over Ap 2-1 (the On position: a beam over the source), blue is the average HI emission taken from four positions separated by a beam from the source in direction of the four galactic cardinal points (the Off position), and the subtraction between them is presented in black, which has a 3$\sigma$ uncertainty of $\sim$ 10 K. The figure shows absorption features at v $\sim$ 12 and 22 km s$^{-1}$, which, taking into account the galactic longitude, may be due to absorption produced by the source against the local gas and the Carina-Sagitarius arm, respectively. Then, a last and conspicuous HI absorption feature is observed at v $\sim$ 33 km s$^{-1}$. By applying the flat galactic rotation curve of Fich et al. (1989), v $\sim$ 33 km s$^{-1}$ gives the possible distances of 2.3 or 11.4 kpc. Taking into account the absence of other absorption features at larger velocities, and mainly at the tangent point (v $\sim$ 93 km s$^{-1}$), following Kolpak et al. (2003), we conclude that Ap 2-1 is located at the near distance of $\sim$ 2.3 kpc. This value is in agreement with the distance estimated for the molecular cloud G35.2-0.74, suggesting a relation between Ap 2-1 and the molecular gas. Hereafter we adopt 2 kpc for Ap 2-1 and the molecular cloud.

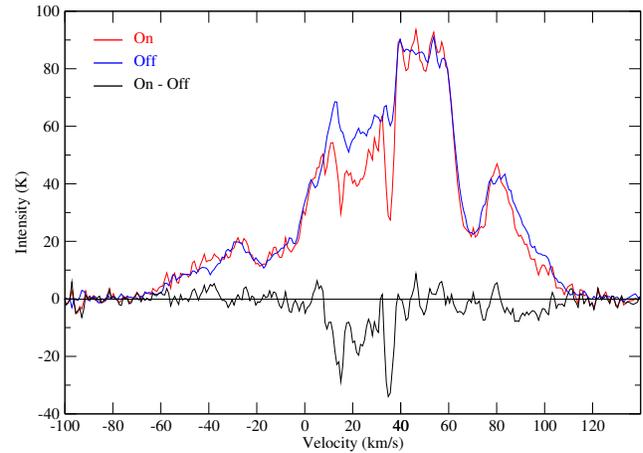

**Figure 1.** HI spectra: the HI emission obtained towards Ap 2-1 (the On position) is the red espectrum, the average HI emission from four positions around the source (the Off position) is displayed in blue, and the subtraction between them is presented in black. The 3$\sigma$ uncertainty of the subtraction is $\sim$ 10 K.

### 3.2 The environment in mid-IR emission

Figure 2 (up) shows a *Spitzer*-IRAC three color image as extracted from GLIMPSE of a region about 20′ × 20′ in the vicinity of the nebula Ap 2-1. The three IR bands are 3.6 $\mu$m (in blue), 4.5 $\mu$m (in green) and 8 $\mu$m (in red). The position of the nebula and the EGO G35.20-0.74 are indicated. As noted by Phillips & Ramos-Larios (2008), Ap 2-1 is enclosed within a larger HII region. Indeed, the nebula apparently lies over the border of an HII region which brights mostly in the 8 $\mu$m emission, mainly originated in the policyclic aromatic hydrocarbons (PAHs). This is because these molecules are destroyed inside the ionized region, but are excited in the photodissociation region (PDR) by the radiation leaking from the HII region (Pomarès et al. 2009). Also the 8 $\mu$m emission shows another HII region towards the galactic southwest. Between the mentioned HII regions there is some diffuse 8 $\mu$m emission, suggesting that they may be linked, i.e. they might be part of the same complex. This possible HII complex is absent in any HII region catalog. Figure 2 (bottom) displays the 20 cm emission obtained from the VGPS (in blue), the MIPS 24 $\mu$m emission (in green) and the 8 $\mu$m emission (in red). Note that MIPS 24 $\mu$m presents blanked pixels towards the EGO and Ap 2-1. The superposition of the three emissions is displayed in white. The 24 $\mu$m emission, not analysed by Phillips & Ramos-Larios (2008), reveals the presence of hot dust. This emission peaks at the position of Ap 2-1 and at the center of the HII region related to the nebula, suggesting that the winds of the stars responsible for the generation of these objects have not yet succeeded in clearing out or destroying all the dust (Watson et al. 2008). On the other hand, the southwestern HII region presents weak emission in this band.

### 3.3 Molecular environment

In this section, using the $^{13}$CO J=1–0 line, we study the molecular cloud G35.2-0.74. As mentioned in Sec. 2.1, this cloud is located at the same distance as Ap 2-1, $\sim$ 2 kpc.

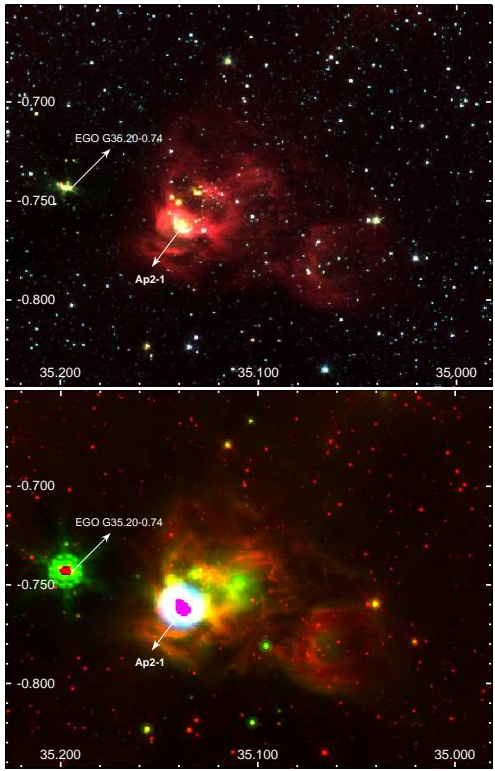

**Figure 2.** *Up*: *Spitzer*-IRAC three color image displayed in galactic coordinates, where the 3.6 $\mu$m emission is presented in blue, the 4.5 $\mu$m in green, and the 8 $\mu$m in red. The position of Ap 2-1 nebula and the EGO G35.20-0.74 are indicated. *Bottom*: Three color image of the same area presented at the upper image. Blue represents the 20 cm emission obtained from the VGPS, green is the MIPS 24 $\mu$m emission, and red is the IRAC 8 $\mu$m. Note that MIPS 24 $\mu$m presents blanked pixels towards the EGO and Ap 2-1.

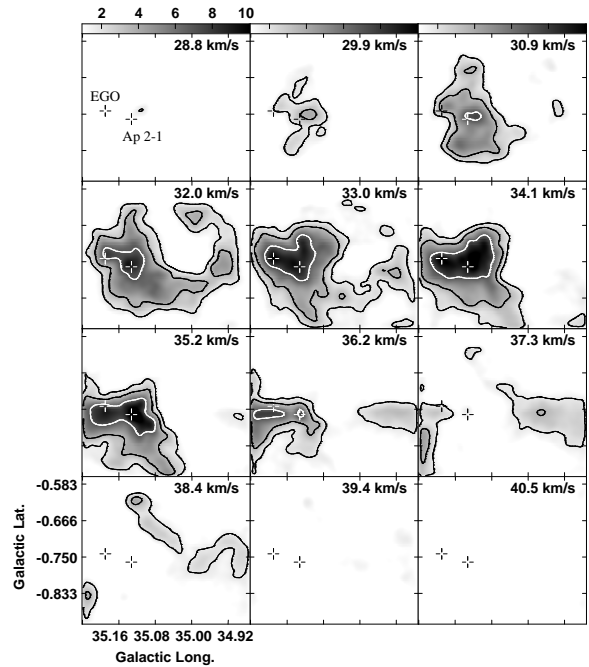

**Figure 3.** Integrated velocity channel maps of the $^{13}$CO J=1–0 emission every $\sim$ 1 km s$^{-1}$. The grayscale is displayed at the top of the figure and is in K km s$^{-1}$, the contour levels are 1.8, 3.5 and 7 K km s$^{-1}$. The position of the EGO and Ap 2-1 are indicated with crosses.

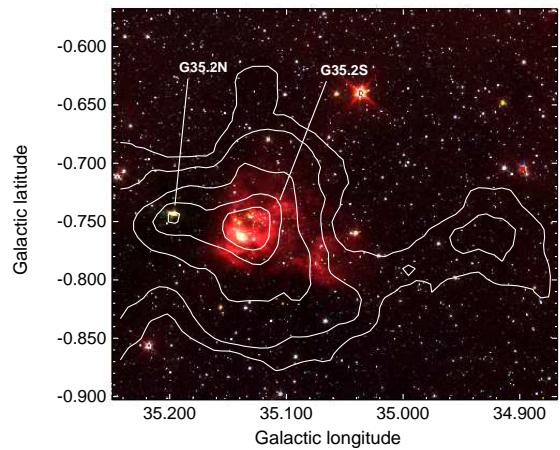

**Figure 4.** IRAC three color image as presented in Fig. 2 (up) with contours of the $^{13}$CO J=1–0 emission integrated between 29 and 39 km s$^{-1}$. The contour levels are 12.5, 18, 30, 42 and 50 K km s$^{-1}$. The location of the molecular condensations G35.2N and G35.2S is indicated.

Figure 3 displays the integrated velocity channel maps of the $^{13}$CO J=1–0 emission every $\sim$ 1 km s$^{-1}$, showing the kinematical and morphological structure of the molecular cloud G35.2-0.74. The position of the EGO and Ap 2-1 are indicated with crosses.

Figure 4 shows the IRAC three color image with contours of the $^{13}$CO J=1–0 emission integrated between 29 and 39 km s$^{-1}$. As indicated in previous works (e.g. Dent et al. 1984; Gibb et al. 2003), the molecular cloud is composed by two main condensations: G35.2N, where the EGO is embedded (Cyganowski et al. 2008), and G35.2S where is located Ap 2-1. The $^{13}$CO emission presented in this work shows that the molecular gas extends towards the galactic west. The southwestern HII region, described in Sec. 3.2, is located between the border of the molecular condensation G35.2S and the mentioned extension.

We use the $^{13}$CO J=1–0 line to have a rough estimate of the molecular mass and density of the region that contains the condensations G35.2N and G35.2S. Assuming LTE, an excitation temperature of 20 K and that the $^{13}$CO emission is optically thin, we obtain an H$_2$ column density of $\sim 1.6 \times 10^{22}$ cm$^{-2}$, and a molecular mass and a density of $\sim 1 \times 10^4$ M$_\odot$ and $\sim 6 \times 10^3$ cm$^{-3}$, respectively. These values are in agreement with those estimated by Little et al. (1983). Of course, the density might be higher in some regions within the molecular cloud, such as G35.2N, where several lines of high density tracers have been detected (Gibb et al. 2003).

## 4 STAR FORMATION

Previous molecular works (see Section 1) suggest that this region, mainly the molecular condensation G35.2N, is active in star formation. In this work, to look for primary tracers of star forming activity, we use the GLIMPSE I Spring'07 Catalog to perform photometry. Considering only sources that have been detected in the four *Spitzer*-IRAC bands, we found 2196 sources in the whole area shown in Fig. 4. The photometric study is presented in Fig. 5. Using the stellar evolutionary stages defined by Allen et al. (2004), we found that 90 sources lie in the region of the protostars with circumstellar envelopes (Class I, red triangles), 140 sources lie in the region of young stars with only disk emission (Class II and intermediate Class I/II, blue triangles) and 1892 sources lie in the region of the main sequence and giant stars (Class III, green triangles). Sources represented as empty black triangles, located outside the delimited regions, could therefore be reddened Class II objects (Allen et al. 2004). In Fig. 6 we present the physical position of Class I and II objects superimposed over the IRAC 8 $\mu$m emission with contours of the integrated $^{13}$CO J=1–0 emission. The figure shows a YSO candidates concentration towards the cloud G35.2N. Also can be appreciated some sources lying towards G35.2S, where the Ap 2-1 nebula and a larger HII region are embedded, and in the area between G35.2S and the southwestern HII region. Within a region with a radius of $\sim 2'$ centered at the position of the EGO G35.20-0.74 in G35.2N there are 37 YSO candidates, while in a region of the same size centered at the position of the HII region related to Ap 2-1 there are only 11 YSO candidates. This difference suggests that the region where is located Ap 2-1 could be older than G35.2N.

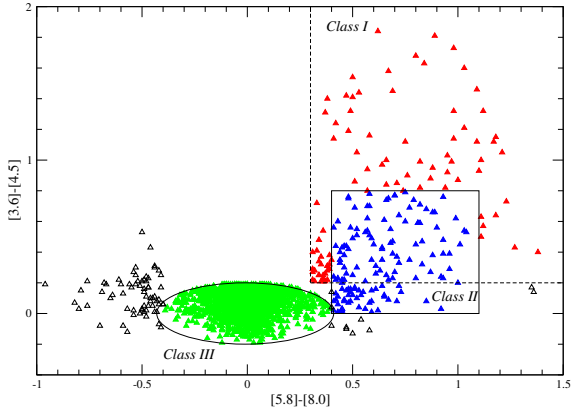

**Figure 5.** GLIMPSE-IRAC color-color diagram [3.6] – [4.5] versus [5.8] – [8.0] for the sources observed in the whole area presented in Fig. 4. Class I, II and III regions are indicated following Allen et al. (2004).

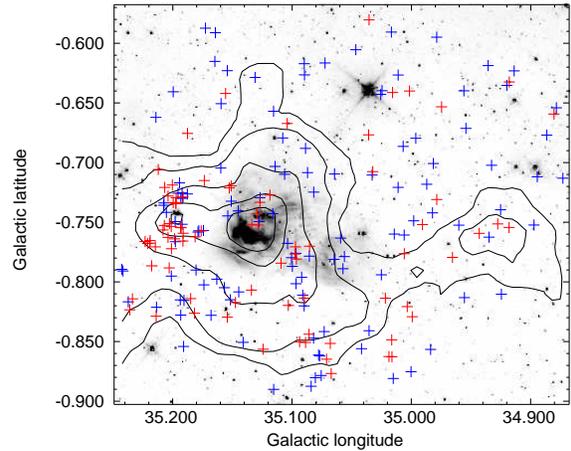

**Figure 6.** *Spitzer*-IRAC 8 $\mu$m emission with YSO candidates superimposed. Red crosses indicate Class I sources and blue crosses are Class II or intermediate Class I/II. The contours correspond to the $^{13}$CO J=1–0 emission as presented in Fig. 4.

In order to study deeper the star formation in the region, we analyse the sources lying near and within the $^{13}$CO contour of 30 K km s$^{-1}$, because they are the most likely embedded sources in the molecular cloud. From the IRAC sources in this region we select those that have detection in at least two bands of the 2MASS *JHK* bands with the best quality. The 40 selected sources are shown in Fig. 7. We perform a fitting of the fluxes of the selected YSO candidates in the IRAC and 2MASS bands to derive their spectral energy distribution (SED) and constrain their evolutionary stage. The SED was obtained using the tool developed by Robitaille et al. (2006, 2007) available online[1]. We use a distance range of 1.5 – 2.5 kpc and a visual absorption range of 10 – 20 mag, which was extracted from the position of the sources in the (*H-K*) versus (*J-H*) color-color diagram (not presented here). This range is in agreement with a visual absorption of Av $\sim$ 18 mag, obtained from Av = $5 \times 10^{-22}$ N(H) (Bohlin et al. 1978), where N(H) = N(HI) + 2N(H$_2$) is the line-of-sight hydrogen column density towards the region. In Table 1 we present the main results of the best-fit model for each source and the number of models which satisfy the following equation:

$$\chi^2 - \chi^2_{min} < 2N_{fluxes},$$

where $\chi^2_{min}$ is the minimum value of the $\chi^2$ (here is the total $\chi^2$, not per data point) among all models, and $N_{fluxes}$ is the number of input data fluxes. Hereafter, the models that satisfies this equation are called *satisfying-models*. The columns of the table are: the source identification (from Fig. 7), $\chi^2_{min}$, central source mass, envelope mass, envelope accretion rate, source age, and the number of satisfying-models, respectively. All cases were checked whether the source cannot be fit simply by a star with interstellar extinction.

To relate the SED to the evolutionary stage of the YSO, Robitaille et al. (2006) defined three different stages as a complement to the "class" classification. These stages are based on the values of the central source mass $M_\star$, the disk mass $M_{disk}$, and the envelope accretion rate $\dot{M}_{env}$ of the YSO. Stage I YSOs are those that have $\dot{M}_{env}/M_\star > 10^{-6}$ yr$^{-1}$, i.e., protostars with large accretion envelopes; stage II are those with $M_{disk}/M_\star > 10^{-6}$ and $\dot{M}_{env}/M_\star < 10^{-6}$ yr$^{-1}$, i.e., young objects with prominent disks; and stage III are those with $M_{disk}/M_\star < 10^{-6}$ and $\dot{M}_{env}/M_\star < 10^{-6}$ yr$^{-1}$, i.e. evolved sources where the flux is dominated by the central source. The stage classification is based on the physical properties of the YSO, rather than on the spectral index derived from the slope of its SED. By inspecting the values

---

[1] http://caravan.astro.wisc.edu/protostars/

Table 1. Parameters of YSO candidates derived from the SED best fit-model.

| Source | $\chi^2_{min}$ | $M_\star$ ($M_\odot$) | $M_{disk}$ ($M_\odot$) | $M_{env}$ ($M_\odot$) | $\dot{M}_{env}$ ($M_\odot$ yr$^{-1}$) | Age (yr) | satisfying-models |
|---|---|---|---|---|---|---|---|
| 1 | 0.97 | 2.39 | $1.65 \times 10^{-3}$ | 4.04 | $2.95 \times 10^{-5}$ | $3.0 \times 10^4$ | 1200 |
| 2 | 2.44 | 4.83 | $3.25 \times 10^{-3}$ | 14.8 | $7.61 \times 10^{-5}$ | $1.2 \times 10^4$ | 58 |
| 3 | 2.47 | 0.18 | $1.38 \times 10^{-2}$ | 0.17 | $4.40 \times 10^{-6}$ | $2.5 \times 10^3$ | 134 |
| 4 | 9.87 | 8.15 | $4.40 \times 10^{-3}$ | 321 | $1.13 \times 10^{-3}$ | $6.5 \times 10^3$ | 515 |
| 5 | 69.9 | 6.36 | $4.83 \times 10^{-3}$ | 481 | $1.80 \times 10^{-4}$ | $9.6 \times 10^4$ | 20 |
| 6 | 104.3 | 3.24 | $1.97 \times 10^{-2}$ | $5.4 \times 10^{-5}$ | 0 | $2.4 \times 10^6$ | 34 |
| 7 | 1.26 | 0.22 | $5.78 \times 10^{-3}$ | $1.8 \times 10^{-2}$ | $3.47 \times 10^{-6}$ | $2.3 \times 10^5$ | 734 |
| 8 | 28.9 | 4.39 | $2.71 \times 10^{-3}$ | 80.7 | $4.84 \times 10^{-4}$ | $8.7 \times 10^4$ | 12 |
| 9 | 3.12 | 2.44 | $1.84 \times 10^{-8}$ | $3.2 \times 10^{-8}$ | 0 | $6.0 \times 10^6$ | 553 |
| 10 | 0.88 | 3.15 | $8.68 \times 10^{-2}$ | 0.24 | $1.08 \times 10^{-5}$ | $1.4 \times 10^5$ | 642 |
| 11 | 6.36 | 9.56 | $3.50 \times 10^{-1}$ | 110 | $1.42 \times 10^{-4}$ | $6.0 \times 10^3$ | 503 |
| 12 | 0.35 | 11.53 | $1.04 \times 10^{-2}$ | 70.6 | $2.13 \times 10^{-4}$ | $2.8 \times 10^4$ | 223 |
| 13 | 1.95 | 5.57 | $2.02 \times 10^{-2}$ | 2.96 | $2.47 \times 10^{-5}$ | $3.0 \times 10^5$ | 1595 |
| 14 | 2.83 | 3.25 | $5.69 \times 10^{-8}$ | $4.7 \times 10^{-4}$ | 0 | $9.4 \times 10^6$ | 218 |
| 15 | 0.09 | 2.33 | $9.83 \times 10^{-3}$ | $7.4 \times 10^{-9}$ | 0 | $2.9 \times 10^6$ | 2927 |
| 16 | 0.53 | 2.13 | $6.68 \times 10^{-4}$ | 0.11 | $2.60 \times 10^{-6}$ | $1.8 \times 10^5$ | 3915 |
| 17 | 1.71 | 7.24 | $1.32 \times 10^{-1}$ | 16.9 | $1.10 \times 10^{-4}$ | $6.6 \times 10^4$ | 1591 |
| 18 | 1.26 | 1.85 | $6.30 \times 10^{-4}$ | 0.11 | $1.40 \times 10^{-5}$ | $1.4 \times 10^5$ | 807 |
| 19 | 2.66 | 1.67 | $7.60 \times 10^{-4}$ | 0.35 | $7.49 \times 10^{-6}$ | $1.3 \times 10^5$ | 5516 |
| 20 | 6.92 | 1.70 | $1.10 \times 10^{-5}$ | $2.0 \times 10^{-7}$ | 0 | $7.8 \times 10^6$ | 1197 |
| 21 | 0.30 | 7.88 | $7.53 \times 10^{-1}$ | 24.2 | $1.89 \times 10^{-4}$ | $7.0 \times 10^3$ | 1941 |
| 22 | 0.28 | 3.81 | $5.67 \times 10^{-2}$ | $7.3 \times 10^{-2}$ | $6.68 \times 10^{-7}$ | $5.0 \times 10^5$ | 2231 |
| 23 | 11.59 | 1.90 | $1.80 \times 10^{-8}$ | $4.7 \times 10^{-9}$ | 0 | $7.9 \times 10^6$ | 646 |
| 24 | 3.28 | 1.52 | $2.28 \times 10^{-3}$ | 1.58 | $1.73 \times 10^{-5}$ | $1.4 \times 10^4$ | 5598 |
| 25 | 11.00 | 2.99 | $6.15 \times 10^{-4}$ | $4.5 \times 10^{-8}$ | 0 | $1.2 \times 10^6$ | 1777 |
| 26 | 0.63 | 2.38 | $1.70 \times 10^{-2}$ | 3.03 | $1.62 \times 10^{-5}$ | $3.5 \times 10^4$ | 420 |
| 27 | 1.46 | 6.71 | $2.92 \times 10^{-1}$ | 8.60 | $6.54 \times 10^{-5}$ | $1.4 \times 10^5$ | 242 |
| 28 | 3.23 | 2.94 | $1.04 \times 10^{-3}$ | 0.14 | $7.34 \times 10^{-6}$ | $3.7 \times 10^5$ | 7079 |
| 29 | 0.55 | 0.51 | $1.02 \times 10^{-2}$ | $3.4 \times 10^{-2}$ | $1.01 \times 10^{-5}$ | $5.4 \times 10^4$ | 841 |
| 30 | 1.83 | 2.57 | $1.44 \times 10^{-3}$ | $1.8 \times 10^{-8}$ | 0 | $5.2 \times 10^6$ | 600 |
| 31 | 23.4 | 3.43 | $1.29 \times 10^{-6}$ | $7.7 \times 10^{-7}$ | 0 | $1.6 \times 10^6$ | 97 |
| 32 | 2.47 | 2.25 | $2.09 \times 10^{-2}$ | $2.4 \times 10^{-7}$ | 0 | $3.8 \times 10^6$ | 414 |
| 33 | 0.22 | 1.31 | $8.69 \times 10^{-4}$ | $3.1 \times 10^{-2}$ | $3.97 \times 10^{-6}$ | $8.9 \times 10^4$ | 780 |
| 34 | 5.05 | 2.31 | $4.18 \times 10^{-5}$ | $1.9 \times 10^{-2}$ | $9.09 \times 10^{-6}$ | $1.9 \times 10^5$ | 166 |
| 35 | 5.21 | 2.08 | $7.29 \times 10^{-4}$ | $2.8 \times 10^{-8}$ | 0 | $8.6 \times 10^6$ | 202 |
| 36 | 0.74 | 1.94 | $2.19 \times 10^{-3}$ | $1.6 \times 10^{-6}$ | 0 | $7.1 \times 10^6$ | 1632 |
| 37 | 2.77 | 3.27 | $1.02 \times 10^{-5}$ | $4.0 \times 10^{-7}$ | 0 | $1.6 \times 10^6$ | 789 |
| 38 | 0.95 | 2.20 | $1.18 \times 10^{-2}$ | 0.47 | $1.87 \times 10^{-6}$ | $3.8 \times 10^5$ | 1934 |
| 39 | 0.31 | 2.19 | $2.26 \times 10^{-6}$ | $1.6 \times 10^{-8}$ | 0 | $7.6 \times 10^6$ | 4943 |
| 40 | 0.25 | 3.83 | $2.19 \times 10^{-4}$ | $1.6 \times 10^{-7}$ | 0 | $4.4 \times 10^6$ | 990 |

presented in Table 1 we found that sources 1, 2, 3, 4, 5, 7, 8, 10, 11, 12, 13, 16, 17, 18, 19, 21, 24, 26, 27, 28, 29, 33, 34, and 38 are stage I, sources 6, 15, 20, 22, 25, 30, 32, 35, 36, and 37 are stage II and sources 9, 15, 23, 31, 39, and 40 are stage III YSOs. It is important to note that most the youngest sources in the region (stage I sources), lie towards G35.2N, conforming a cluster of YSOs (see Fig. 7). The presence of this cluster suggests an age gradient in the physical distribution of the sources: the youngest YSOs are embedded in G35.2N, while the evolved YSOs are embedded preferently in G35.2S and in its surroundings. As pointed by Robitaille et al. (2006), a way to go deeper in identifying young objects from the SED is looking for sources that can be fitted only by satisfying-models with $\dot{M}_{env} > 0$. Sources that are fitted by satisfying-models with zero and non-zero values of $\dot{M}_{env}$ may be more evolved YSOs whose evolutionary stage could be not well constrained by the data (see Poulton et al. 2008). Only sources 2 and 5, both embedded in G35.2N, have $\dot{M}_{env} > 0$ for all of their satisfying-models.

Of course the evolutionary stage of the sources cannot be assured without doubt because we are comparing only the values of the best fit-model. Given to the large number of satisfying-models (see last column in Table 1), the parameters can be spread over a wide range, and thus they not necessarily provide a good estimate of the evolutionary stage. To improve this, it will be necessary to include fluxes at other wavelengths, mainly at far-IR and sub-mm, to reduce the number of satisfying-models.

Finally we derive the SED for the EGO G35.20-0.74 using the fluxes in the *JHK* 2MASS bands (*J* band flux is an upper limit), the four IRAC bands, the IRAS 25, 60 and 100 $\mu m$ bands, and the 450 and 850 $\mu m$ bands from SCUBA (Di Francesco et al. 2008). Figure 8 shows the YSO SED fits (up), and the best fit-model showing the total flux and the fluxes from the envelope and the disk (bottom). The physical parameters from the best fit-model are: $M_\star = 15.1$ $M_\odot$, $M_{disk} = 0.41$ $M_\odot$, $M_{env} = 3.8 \times 10^3$ $M_\odot$, and $\dot{M}_{env} = 5.3 \times 10^{-3}$ $M_\odot$ yr$^{-1}$. A large amount of mod-els have $\dot{M}_{env}/M_\star > 10^{-6}$ yr$^{-1}$, which indicates that the EGO is likely a massive protostar(s) with a large accretion envelope. These results are in agreement with previous high resolution studies performed in the region. Gibb et al. (2003) discovered, in an area of $20'' \times 20''$ centered at R.A.= $18^h58^m13^s$, DEC. = + $01°40'36''$ J2000.0 ($l = 35.197$, $b = -0.742$), a cluster of massive outflow sources. The authors propose that one of them, the source G35MM2, is a massive YSO still in the collapse phase with a surrounding HII region that is quenched by the action of the infalling gas. Previously, Fuller et al. (2001) suggested that the observed jets in the region arise from a disk around a B1 star with a mass of $M_{disk} \sim 0.15$ M$_\odot$, while Dent et al. (1985) derived a spectral class of B0.5 for the exciting star, and estimated a stellar mass of $\sim 15$ M$_\odot$.

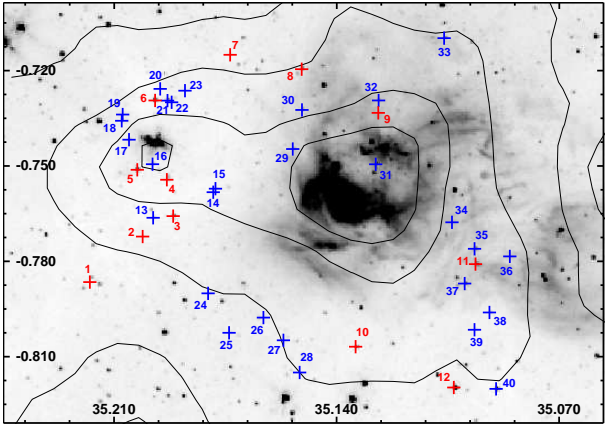

**Figure 7.** *Spitzer*-IRAC 8 µm with contours of the $^{13}$CO integrated emission. The crosses are the selected sources, i.e. sources with detection in the four IRAC bands and at least in two 2MASS bands with the best quality, near and within the $^{13}$CO contour of 30 K km s$^{-1}$. Red and blue crosses indicate Class I and II sources, respectively.

We conclude that this region, mainly G35.2N, where we discovered a cluster of YSOs, is indeed active in star formation. From the SED analysis we obtained a first glimpse about the physical parameters of the YSOs in the region. Concerning to Ap 2-1 nebula, there seems little doubt that it is associated with the star forming complex. In what follows, using optical data we analyse specifically this nebula.

## 5   AP 2-1 NEBULA

### 5.1   Observations and data reduction

The optical observations were carried out on August 10, 2007, using the REOSC spectrograph attached to the 2.15 m telescope at CASLEO, Argentina. The spectra were taken with a 300 line mm$^{-1}$ grating, which provides a dispersion of 3.4 Å px$^{-1}$, and a wavelength range of 3500–7000 Å. In addition we analyse better resolution (0.61 Å px$^{-1}$) spectra obtained from observations carried out on July 22 and 23, 2004 (Weidmann & Carranza 2005). The slit was always centered on the nebula ionizing star, at different position angles, and opened to $3''$ (consistent with the seeing at the site). We obtained three spectra with an exposure time of 3600 s. The optical spectra were reduced using IRAF[2], following standard techniques.

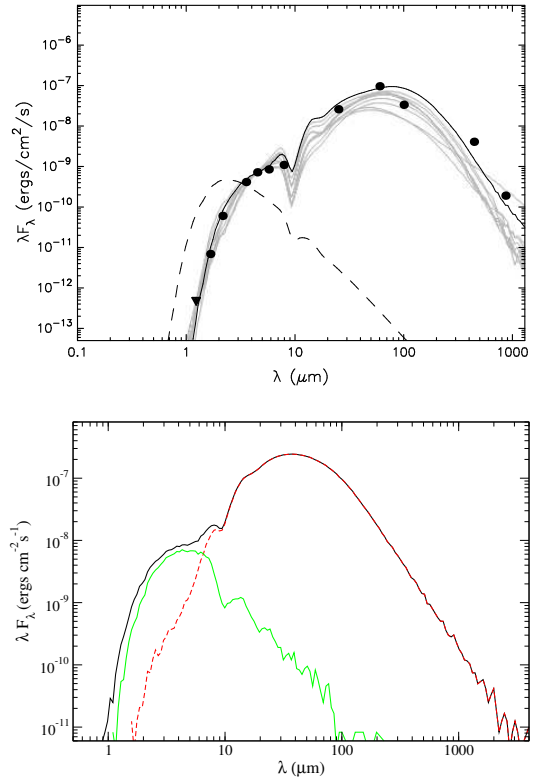

**Figure 8.** SED for the EGO G35.20-0.74. Up: black line shows the best fit, and the gray lines show subsequent good fits. The dashed line shows the stellar photosphere corresponding to the central source of the best fitting model, as it would look in the absence of circumstellar dust. The points are the input fluxes. Bottom: best fit-model showing the total flux (black line), the envelope flux (red dashed line) and the disk flux (green line). Note that the main contribution to the total flux comes from the envelope.

### 5.2   The emission nebula

Figure 9 displays an optical spectrum of Ap 2-1. It is important to remark the absence of [OIII] lines at $\lambda\lambda$ 4959, 5007Å in the spectra. They are usually the strongest emission lines in the optical spectra of a PN.

Using the $H\alpha/H\beta$ ratio we derive the interstellar extinction c($H_\beta$). Assuming a temperature of $10^4$ K and an electron density of $10^2$ cm$^{-3}$, we use the $H\alpha/H\beta$ intrinsic value of 2.86 (case B of recombination; Brocklehurst 1971). Using the standard extinction law of Seaton (1979), we obtain c($H\beta$) = 2.76 (Av = 5.8). This interstellar extinction is likely due to the molecular cloud associated with Ap 2-1. However, Av = 5.8 is lower than the visual absorption range considered in Section 4. Moreover, the extended HII region and the southern shell source where Ap 2-1 is located are visible in $H\alpha$. Figure 10 shows the 8 $\mu$m and $H\alpha$ emission in red and green, respectively. The superposition of both emissions is displayed in yellow. The continuum-subtracted $H\alpha$ emission was extracted from the INT/WFC Photometric $H\alpha$ Survey (IPHAS)[3]. Thus, we suggest that the HII region complex is not highly embedded, and probably is located at the nearest border of the molecular cloud.

---

[2] IRAF: the Image Reduction and Analysis Facility is distributed by the National Optical Astronomy Observatories, which is operated by the Association of Universities for Research in Astronomy, Inc. (AURA) under cooperative agreement with the National Science Foundation (NSF).

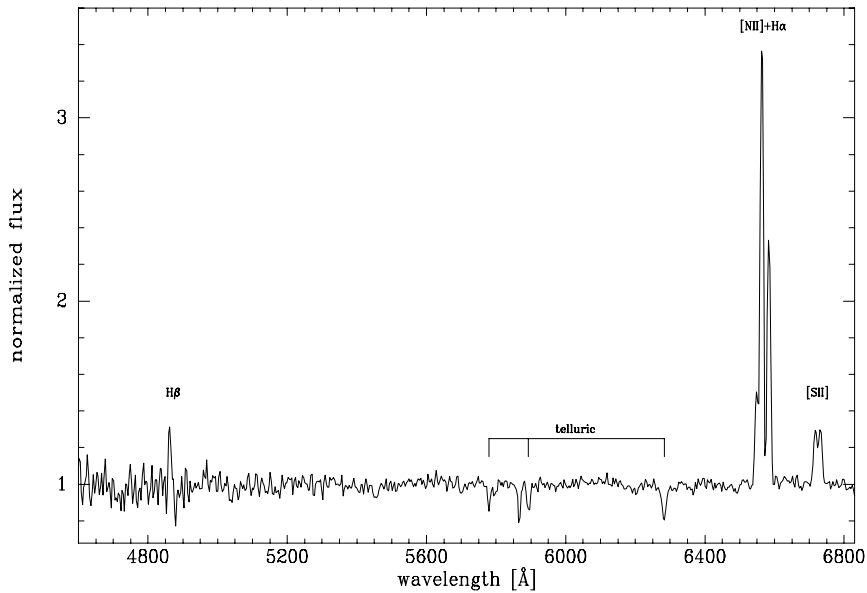

**Figure 9.** Optical low resolution spectrum of Ap 2-1. The detected emission lines are indicated.

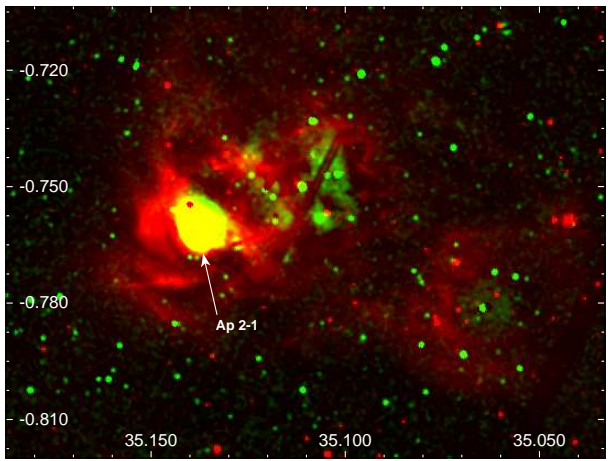

**Figure 10.** Spitzer 8 $\mu$m (in red) and IPHAS $H\alpha$ (in green) emission around Ap 2-1 Nebula. The superposition of both emissions is displayed in yellow.

The highest resolution spectra allowed us to determine the radial velocity of the nebula in $v_{LSR} = 31 \pm 16$ km s$^{-1}$, in agreement with the velocity of the HI absorption (see Sec. 3.1) and the molecular gas (see Sec. 3.3). This value was obtained from the $H\alpha$ and [NII] lines, and the uncertainty was calculated following Keel (1996). On the other hand, the line width of the nebular gas (FWHM) is comparable to the width of the arc-lamp lines, suggesting a low expansion velocity or that the nebula is static respect to the surrounding ISM.

### 5.3 The ionizing star

Although the stellar continuum shows a reasonable S/N ratio, neither absorption nor emission feature is observed. The peak of the dereddened continuum spectrum of the Ap 2-1 central star is bluer than 4100 Å. By fitting a blackbody function to this spectrum we obtain a temperature of $\sim 28000$ K for the star. Such a temperature is compatible with an early star. HII regions and PNe with significant [OIII] emission are ionized by stars with $T_{eff} > 35000$ K (Stasińska 1990). Thus, the ionizing source of Ap 2-1, that does not present [OIII] emission, might have an effective temperature of 20000 K $<$ T$_{eff}$ $<$ 35000 K. The lower limit is due to that it is necessary a $T_{eff} > 20000$ K to ionize the hydrogen.

Photometric measurements of the central star are generally contaminated by the emission of the nebula. This problem can be overcome by spectrophotometric measurements. Thus, after manual removal of the nebular emission lines, we integrated the spectrum over the response functions of the standard B and V photometric bands, and we obtain the magnitudes B = 16.1 and V = 15.1 for the core. The V magnitude is comparable with that obtained by Tylenda et al. (1991), while our B magnitude is significantly lower than the reported by the authors.

We performed a further test to have a rough value of the absolute magnitude of the central star. Assuming a distance of 2 kpc and using Av = 5.8, we obtain Mv = $-2.1$ for the star, suggesting a B2V star (Walborn 1972). Adopting a bolometric correction of BC= $-2.36$ (Hayes 1978), we obtain a luminosity of L = 4800 L$_\odot$ for the star. On the other hand, we estimate the number of UV photons nec-essary to keep the gas ionized using $N_{UV}(\text{photons s}^{-1}) = 0.76 \times 10^{47} \, T_4^{-0.45} \, \nu_{GHz}^{0.1} \, S_\nu \, D_{kpc}^2$ (Chaisson 1976), where $T_4$ is the electron temperature in units of $10^4$ K, $D_{kpc}$ is the distance in kpc and $S_\nu$ is the flux density in Jy. Adopting an electron temperature of $10^4$ K, a distance of 2 kpc, and using $S_{30GHz} = 241$ mJy (Pazderska et al. 2009) we obtain $N_{UV} \sim 1 \times 10^{47}$ s$^{-1}$, which roughly agrees with the density of lyman-continuum photons emitted by an early B-type star (Vacca et al. 1996; Alvarez et al. 2004). Additionally, using the HK 2MASS magnitudes of the Ap 2-1 ionizing star ($H = 9.879 \pm 0.027$ mag, and $K = 9.502 \pm 0.026$ mag), we observe that the source position in a ($H$) versus ($H$-$K$) color-magnitude diagram suggests that its spectral type could lie between B0V and B1V. These results are consistent with an scenario of an HII region around a main sequence star. The spectral type of the Ap 2-1 ionizing star is comparable with that of the ionizing star of the HII region embedded in G35.2N (Fuller et al. 2001), suggesting that the region is not only rich in star formation, but also probably in massive stars such as B-type stars.

---

[3] http://www.iphas.org/index.shtml

## 6   SUMMARY AND CONCLUSIONS

Using multiwavelength surveys and archival data we studied the molecular cloud G35.2-0.74, where the Ap 2-1 nebula is embedded. Additionally, the nebula was studied using optical data obtained with the 2.15 m telescope at CASLEO, Argentina. The main results can be summarised as follows:

(a) From the HI absorption study we estimate a distance of $\sim 2$ kpc for Ap 2-1 in coincidence with the molecular cloud G35.2-0.74.

(b) We find that Ap 2-1 is likely embedded in the southern portion of the molecular cloud G35.2-0.74, being part of an HII region complex with presence of hot dust.

(c) We confirm that the molecular cloud G35.2-0.74 is very active in star formation, mainly towards the north portion (G35.2N), where we discover a cluster of YSOs. From the spectral energy distribution (SED) of the sources, we obtain the physical parameters of most of these YSOs.

(d) From the $H\alpha$ and [NII] lines we obtain a radial velocity of $v_{LSR} \sim 31$ km s$^{-1}$ for Ap 2-1, in agreement with the velocity of the molecular gas. This result confirms that Ap 2-1 is associated with the molecular cloud G35.2-0.74.

(e) Ap 2-1 presents an extinction of $c(H\beta) = 2.76$ (Av = 5.8). [OIII] emission lines are absent from its optical spectra, suggesting low excitation.

(f) We obtain an absolute magnitude and an amount of lyman-continuum photons for the Ap 2-1 ionizing star of Mv $= -2.1$ and $N_{UV} \sim 1 \times 10^{47}$ s$^{-1}$, respectively. These results suggest an early B-type star, which is comparable with the ionizing star of the HII region located at G35.2N.

In the literature there is a controversy about the nature of Ap 2-1. In some works, this nebula was classified as a PN and in others as an HII region. Taking into account our results, we conclude that Ap 2-1 is an HII region probably excited by an early B-type star embedded in a molecular cloud very active in star formation.

## ACKNOWLEDGMENTS

We wish to thank the anonymous referee whose comments and suggestions have helped to improve the paper. S.P. is member of the *Carrera del investigador científico* of CONICET, Argentina. W.W. is a postdoctoral fellow of CONICET, Argentina. S.P and W.W. thank Dr. M. Gómez for some useful corrections in the manuscript. This work was partially supported by Argentina grants awarded by UBA, CONICET and ANPCYT. The CCD and data acquisition system at CASLEO has been financed by R. M. Rich trough U.S. NSF grant AST-90-15827.